\documentclass[twocolumn]{aastex63}
\received{January 28, 2021}
\revised{April 28, 2021}
\accepted{May 11, 2021}

\shorttitle{ALMA Host galaxy observation of the off-axis Gamma-Ray Burst XRF 020903}

\begin{document}

\title{ALMA Host galaxy observation of the off-axis Gamma-Ray Burst XRF 020903}


\author{Jheng-Cyun Chen}
\affiliation{Institute of Astronomy, National Central University, Chung-Li 32054, Taiwan}

\author{Yuji Urata}
\affiliation{Institute of Astronomy, National Central University, Chung-Li 32054, Taiwan}

\author{Kuiyun Huang}
\affiliation{Center for General Education, Chung Yuan Christian University, Taoyuan 32023, Taiwan}

\begin{abstract}

We investigated the radio properties of the host galaxy of X-ray flash, XRF 020903, which is the best example for investigating of the off-axis origin of gamma-ray bursts (GRBs). 
Dust continuum at 233 GHz and CO are observed using the Atacama Large millimeter/submillimeter array.
The molecular gas mass derived by applying the metallicity-dependent CO-to-H$_{2}$ conversion factor matches the global trend along the redshift and stellar mass of the GRB host galaxies. 
The estimated gas depletion timescale (pertaining to the potential critical characteristics of GRB host galaxies) is equivalent to those of GRBs and super-luminous supernova hosts in the same redshift range. 
These properties of the XRF 020903 host galaxy observed in radio resemble those of GRB host galaxies, thereby supporting the identical origin of XRF 020903 and GRBs.
   
\end{abstract}
%


\section{Introduction} \label{sec:intro}
X-ray flashes (XRFs) were discovered by the BeppoSAX satellite during the dawn of the gamma-ray bursts \citep[GRBs;][]{ heise2001, kippen2002}. The observed X-ray properties in the prompt phase were identical to those of GRBs, except for the considerably lower energy values of the spectral peak energy, $E_{\rm peak}$ in the $\nu F_{\nu}$ spectrum. Systematic prompt  phase observations made by HETE-2 indicated that the XRFs are associated with the same phenomenon as classical hard GRBs and are representative of the extension of the GRB population to low-peak-energy events \citep{sakamoto2005}. Three models have been proposed to explain the observed prompt properties: high-redshift origin \citep{heise2003}, intrinsic property \citep[e.g., a subenergetic or inefficient fireball;][]{huang2002}, and off-axis jet models \citep{yamazaki2002,zhang2004,lamb2005}. For the latter two models, on-axis and off-axis orphan GRB afterglows can appear because of their lower Lorentz factor caused by inefficient fireball and off-axis viewing, respectively.  
\citet{urata2015} confirmed the off-axis origin of the XRF 020903 case by identifying achromatic rebrightening in the afterglow. Additionally, XRF 020903 was characterized well through observations, e.g., redshift measurement \citep{soderberg2004}, a lower intrinsic $E^{\rm src}_{\rm peak}$\citep{sakamoto2004}, and supernova association\citep{soderberg2005}.
Among all the XRF samples, XRF 020903 was the only sample suitable for investigating its origin.
Although the XRF observations have been terminated owing to the lack of wide-field soft-X-ray monitoring instruments such as WXM on board HETE-2 \citep{shirasaki2003}, planned GRB missions, i.e., SVOM \citep{svom}, HiZ-GUNDAM\citep{yonetoku2020}, and THESEUS\citep{amati2018} would provide large samples.  

The off-axis viewing of GRB jets for both long and short GRBs is essential for
understanding the unification of GRBs, including related
energetic stellar explosions, as well as to perform multimessenger astronomy.
In particular, short GRBs and the jet viewing angle are essential for
understanding the characteristics of short GRBs associated with the gravitational
wave transients caused by neutron star mergers \citep[e.g.,][]{alexander17, haggard17,lazzati17, murguia17,ioka18, jin18, kathirgamaraju18, troja18, troja19, 2018MNRAS.478..733L, 2018NatAs...2..751L, 2019ApJ...870L..15L}.
Long GRBs are believed to occur when a very massive star dies in a highly energetic supernova (SNe), forming a black hole and producing a relativistic jet. 
The averaged jet opening angles were measured to be $\sim3^{\circ}.5$, i.e., of the same order as that of AGNs, based on achromatic temporal breaks in afterglow light curves \citep[e.g.,][]{racusin09}. Cocoon structures surrounding the GRB jets were identified to be similar to those of AGNs \citep{izzo19, chen20}.
Similar to the unification model of AGNs, observing GRB jets at the various viewing angles can provide GRBs, XRFs, off-axis orphan GRB afterglows, and energetic stellar explosions \citep[e.g.,][]{slsn, huang19, izzo20}.
In this regard, all of these phenomena share the same environment (i.e., common host galaxy properties).

Off-axis orphan GRB afterglow searches were performed at various wavelengths \citep{grindlay1999, greiner2000, rau2006, levinson2002, galyam2006, huang2020}.
Although these numerous surveys do not involve the detection of off-axis orphan GRB afterglows, the detection rate implies that the consideration of jet structures is essential for future orphan GRB afterglow surveys \citep[e.g.,][]{huang2020}.
The VLA Sky Survey (VLASS) identified a luminous radio transient and reported it as a candidate for off-axis orphan GRB afterglows \citep{Law18, Marcote19}. Although the possibility of the nebula of a newly born magnetar is not precluded, the similar star formation property of the host galaxy to that of long GRB hosts supports the off-axis orphan GRB afterglow phenomenon.   

Millimeter and submillimeter observations for both GRBs and their host galaxies using the Atacama Large Millimeter/submillimeter Array (ALMA) have provided new insights. 
The first measurement of the radio linear polarization suggested depolarization caused by nonenergetic electrons, which afforded acceleration efficiency at the shock \citep{urata19}.
Meanwhile, carbon monoxide (CO) lines \citep{hatsukade14, co-grb980425} were identified from spatially resolved imaging observation.  \citet{hatsukade14} indicated that the bursts occurred in regions rich in dust, but not particularly rich in molecular gas.
\citet{co-grb980425} revealed that the presence of starburst modes of star formation on local scales in the galaxy, even though the galaxy as a whole cannot be categorized as a starburst based on its global properties.
By expanding the previous studies using nearby GRB samples \citep[$z<$0.12;][]{michalowski18}, statistical studies regarding CO observations indicated possible common properties of GRB host galaxies in terms of their molecular gas mass fraction and gas depletion timescale, particularly at $z<1$ \citep{michalowski18, hatsukade20}.
Additionally, the detection of [$\rm C_{II}$] also provided a new physical property for the characterization of GRB host galaxies at $z\sim1-2$ \citep{hashimoto19}.

Herein, we present the results of ALMA continuum and CO line observations on the host galaxy of an off-axis GRB event (i.e., XRF 020903).
The remainder of this paper proceeds as follows. In Section 2, we summarize the previous XRF 020903 observations.
In Section 3, we describe the ALMA observations for XRF 020903 host galaxies and other GRB samples. In Sections 4 and 5, we present the results of the continuum and CO line observations and discuss their properties, respectively, based on using the following cosmological parameters: $H_{0}$=70 $\rm {km\ s^{-1}}$, $\Omega_{0}=0.3$, and $\Omega_{\lambda}=0.7$.

\section{Off-Axis GRB, X-Ray Flash 020903}

XRF 020903 is the most favorable sample for the unification of GRBs, whereas the jet viewing angle is the most favorable parameter. 
The event was characterized by (a) measurements of lower spectral peak energy of the prompt emission, (b) redshift measurements ($z=0.251)$, (c) evidence of the off-axis viewing of the GRB jet, (d) association of a supernova component in the optical afterglow, and (e) a low-metallicity environment via an optical identification of the host galaxy.
Based on observation by the HETE-2 satellite, the lowest intrinsic
spectral peak energy $E^{\rm src}_{\rm peak}$ of 3.3$^{+1.8}_{-1.0}$ keV was identified among all the
XRF samples \citep{sakamoto2004}. The redshift of $z=0.251$ \citep{soderberg2004} was measured via optical spectroscopic observation observations. Achromatic rebrightening caused by the off-axis viewing of classical GRB jets \citep{urata2015} were detected via early multicolor optical observations.
SN1998bw-like supernova association was confirmed via optical spectroscopy at $\sim$25 day after the burst\citep{soderberg2005}.

\begin{figure}
\plotone{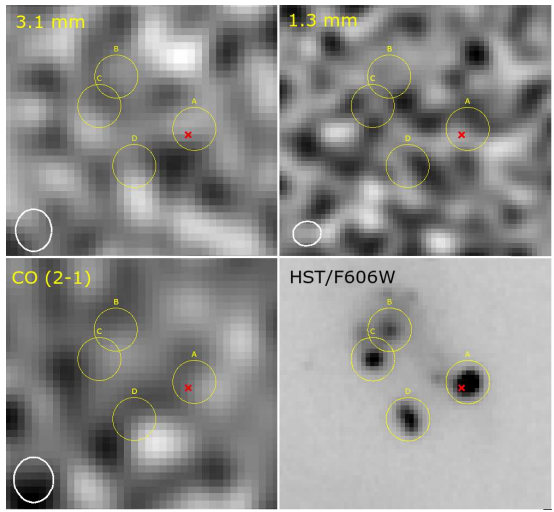}
\caption{Continuum maps at 3.1 mm (top left) and 1.3 mm (top right), CO velocity-integrated maps (bottom left), and optical image (bottom right). The yellow circles with $0".2$ radius indicate the positions of optical components. Afterglow position is marked by the red cross. Synthesized beam size is shown in the lower left corners.\label{image}}
\end{figure}

The high-resolution host galaxy image obtained by the Hubble Space Telescope \citep[HST;][]{soderberg2004} shows the existence of at least four components as shown in Figure \ref{image}. 
The optical spectroscopy for the burst site measured the metallicities as log(O/H)+ 12$\sim$ 8.0.
Combined with GRB host galaxy measurements, the observed metallicities indicated the lowest values \citep{levesque2010}. Meanwhile, burst site studies involving optical spectra also indicated features of a significant Wolf$-$Rayet star population \citep{hammer2006, han2010}.
The spatially resolved optical spectroscopic observations indicated that the bust site is likely the most recent and active site of star formation in the XRF 020903 host galaxy, which is consistent with other spatially resolved GRB host galaxy studies \citep{thorp2018}.
Using line diagnostics \citep{thorp2018}, the low metallicities for all of four regions were measured to be $8.1\pm0.1$ for the burst site and $\sim$8.2 for the remaining regions.  

\section{Observations and Data} \label{sec:events}

\subsection{XRF 020903}

The location of the XRF 020903 field at Band 3 and Band 6, for the Cycle 3 program (Project code: 2015.1.01254.S) was observed using ALMA.
%
The observations in Band 3 were executed on 2016 September 19 and 21. The total on-source time was 4318 s.
A correlator was used in the frequency domain mode with a bandwidth
of 1875 MHz (488.28 kHz $\times$ 3840 channels). 
Four basebands were used, resulting in a total bandwidth of 7.5GHz.
The main aspect of the observation was the search for the CO ($J=1-0$)
line associated with the XRF 020903 host galaxy.
The bandpass and flux were calibrated using observations from J2258-2758
and J0006-0623, and those from J2236-1433 were used for the phase calibration.
%
The photometric observation in Band 6 with the total on-source time of 962 s was performed on 2016 July 28. 
The bandpass and flux were calibrated using observations from J2258-2758
and Pallas, whereas observations from J2303-1841 were used for phase calibration.

\begin{figure}
\epsscale{1.0}
\plotone{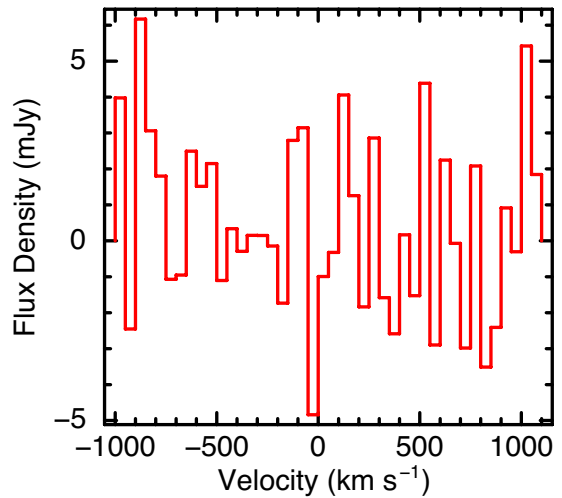}
\caption{CO spectra with the velocity resolution of 47.7 km s$^{-1}$ for the burst site of the XRF 020903.\label{spec}}
\end{figure}

The data were reduced using the Common Astronomy Software Applications package \citep{casa} in a standard manner. Data calibration was performed using the ALMA Science Pipeline Software of CASA version 4.7.0-1.
Maps were processed using the TCLEAN algorithm with the Briggs
weighting (with a robust parameter of 0.5).
Furthermore, a CO ($J=1-0$) velocity-integrated map with a velocity width of 180 km $^{-1}$ was created using the TCLEAN algorithm.
The final synthesized beam size (FWHM) was $0".38 \times 0".33$ for Band 3, $0".27\times0".23$ for Band 6, and $0".42\times0".36$ for the velocity-integrated maps.
These maps, with their optical images obtained using the HST, are shown in Figure \ref{image}.
By resolving the four optical host galaxy components, the 3$\sigma$
upper limits of 0.032 mJy at 113 GHz and 0.061 mJy at 224 GHz were
estimated. The 3$\sigma$ limit of the CO velocity-integrated map with
180 km/s was 0.57 mJy.
The spectra around the CO($J=1-0$) line for the host region was centered at R.A.
= $22^{h}48^{m}42^{s}.37$, decl. = $-20^{\circ}46^{'}08".72$ with a radius of $0".8$(i.e., including all four optical components) indicates a null detection
(Figure \ref{spec}).

\begin{figure*}
\plotone{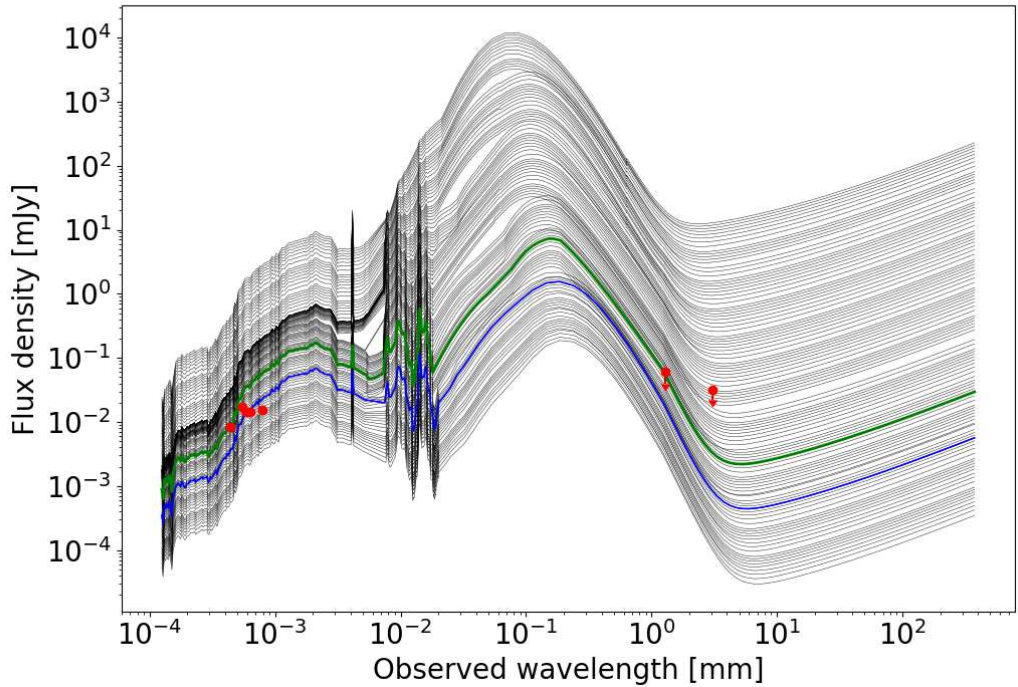}
\caption{SED of the XRF 020903 host galaxy. The ALMA observation constrains the IR SEDs using template scaling. Templates lower than the green solid line satisfy the ALMA 3$\sigma$ upper limit for the XRF 020903 host galaxy. The solid blue line is the closest to the optical measurement based on the high-resolution image captured using HST.}\label{020903temp}
\end{figure*}

\begin{figure*}
\plotone{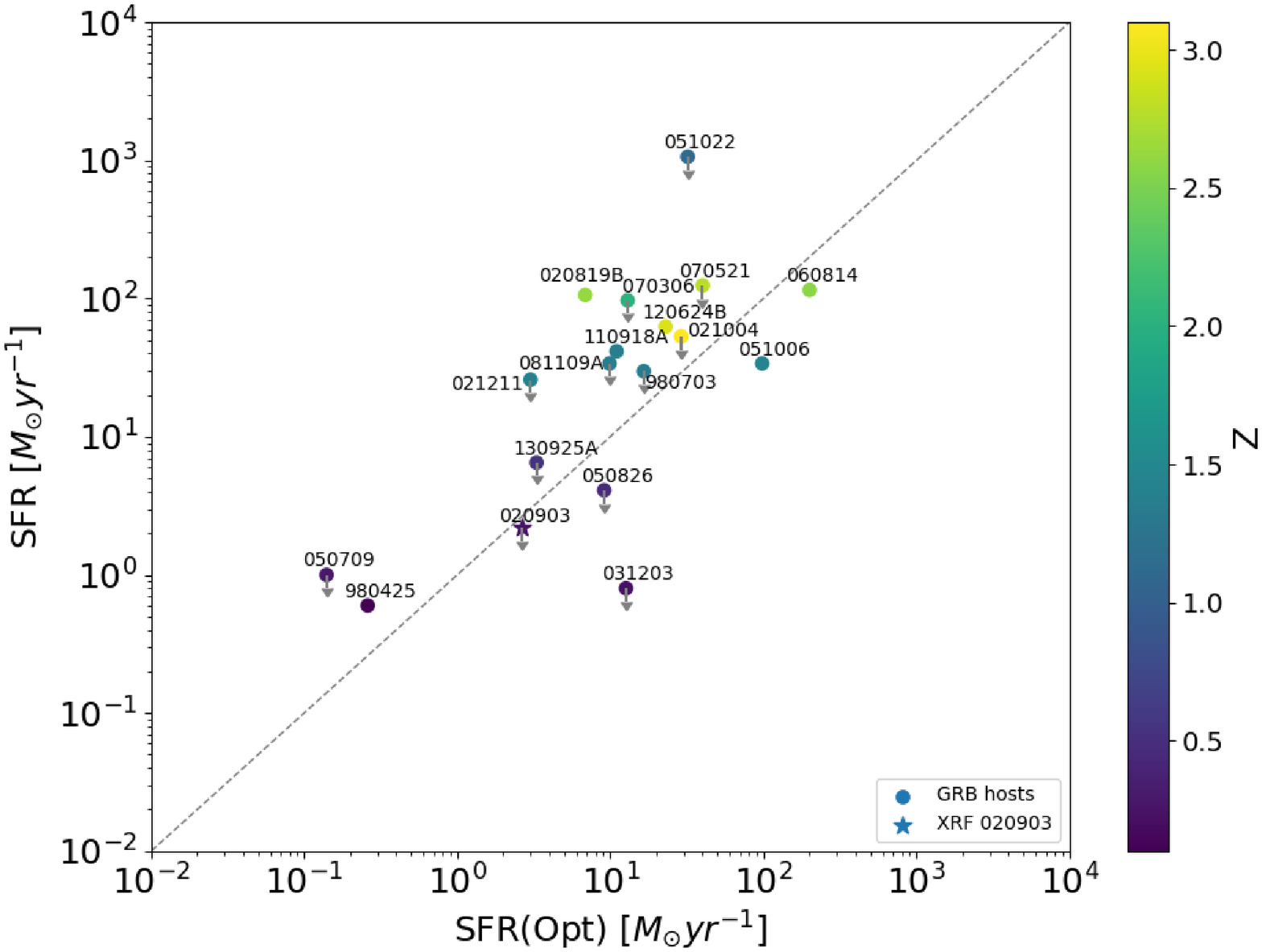}
\caption{The SFRs obtained using the SED template scaling method compared with the measurements obtained using the optical method \citep[i.e., emission lines or UV luminosity;][]{ghost}. 
\label{sfr}}
\end{figure*}

\subsection{GRB Host Galaxy Samples}

Table \ref{tab} summarizes the properties of 36 GRBs and two super-luminous supernova (SLSN) host galaxies observed using the ALMA in comparison with the XRF 020903 measurements. 
These samples included various GRBs, including one short GRB \citep[GRB 050709;][]{050709r1,050709r2,050709r3} and one ultra-long GRB \citep[GRB 130925A;][]{130925r1, 130925r2, 130925r3}.
GRB 050709 is a classical short/hard GRB, but the optical properties of its host galaxy indicate star formation activity \citep{covino06}. 
GRB 130925A is also classified as an optically dark GRB because of its high visual extinction \citep{greiner14}.
The revised redshift \citep{perley17} is used for one of the optically dark GRBs, GRB 020819B for evaluating the observation targeted on the CO line.
Two samples (GRB 090423 and GRB 130606A) may have occurred at the reionization epoch \citep[e.g.,][]{tanvir09, totani14, totani16}.
The neutral fraction of the IGM measured using GRB 130606A implied the incompletion of reionization at $z\sim6$ \citep{totani14, totani16}. This higher redshift event also obeys the energy correlations established using low-redshift GRBs \citep{yasuda17}.
Furthermore, we added two SLSNs (i.e., SN2017egm and PTF10tpz), as potential samples of identical stellar explosions by observing the GRB jet at an off-axis viewing angle \citep[e.g.,][]{nicholl17, wheeler17, coppejans18}. 

These physical properties of the reference samples were obtained by reducing the ALMA archive data and from the literature.
We used the products obtained from the ALMA archive for continuum observations, except for GRB 050709.
These archive data were calibrated and imaged by the ALMA Regional Centers using standard procedures.
Two execution blocks for GRB 050709 observations (Project code: 2016.1.01333.S) at the identical band (Band 6) were reduced separately. We reproduced the calibrated data, merged them, and then imaged them using TCLEAN. 
The total flux measurements and the RMS noise estimations for the nondetection cases were
performed using function "imfit" and "imstat", respectively.
We confirmed that our continuum measurements were consistent with previous results
\citep{wang12, huang17, berger14,  hatsukade20}.  
For Table \ref{tab}, we referred to the detailed measurements made by \citet{endo07}, \citet{hatsukade11}, \citet{wang12}, \citet{berger14}, \citet{hatsukade14}, \citet{stanway15}, \citet{huang17}, \citet{michalowski18}, \citet{ptf10tpz}, \citet{hashimoto19}, \citet{hatsukade19}, \citet{postigo20}, \citet{hatsukade20}, and \citet{hatsukade20egm}.
%

\section{IR Luminosity and Star Formation Rate (SFR) with GRB Host Galaxies}
\label{lifsfr}

The ALMA continuum observations provided unique upper limits for estimating the rest-frame IR luminosity and SFR of GRB and XRF host galaxies. 
Based on \citet{wang12}, a redshift to a template IR SED was applied to these results. 
We referred to the SED library provided by \citet{chary01}, which is luminosity-dependent ($2\times10^8$ to $4\times10^{13} L_\sun$) based on a locally calibrated luminosity-dust temperature relation and does not enable SED scaling.
As shown in Figure \ref{020903temp}, we constrained the 3$\sigma$ upper limits of IR luminosity for the XRF 020903 host galaxy with the Band 6 observation as $L_{\rm IR} <1.3\times10^{10} L_\sun$. 
Moreover, the templates with $L_{\rm IR} <1.3\times10^{10} L_\sun$ are also consistent with optical flux measurements
(the blue line in the Figure \ref{020903temp}). In terms of the SED for describing optical measurements, we primarily used the HST measurement \citep{soderberg2004}, because the ground-based multicolor measurements were affected by the nearby galaxies, as reported by \citet{bersier06}. 
Using the SFR conversion of star-forming galaxies, SFR($M_\sun$ yr$^{-1})=1.7\times10^{-10} L_{\rm IR}/L_\sun$ \citep{kennicutt98}, we obtained the 3$\sigma$ upper limit of the SFR as $<2.2$ ($M_\sun$~yr$^{-1}$). This estimation is consistent with the measurement (2.65$M_{\sun}$ yr$^{-1}$) using $H_{\alpha}$ or rather small \citep{ghost}.

Similarly, we derived the $L_{\rm IR}$ and SFR for reference GRB samples (Table \ref{tab}).
Figure \ref{sfr} shows a comparison between the estimated SFRs and those measured based on $H_{\alpha}$, [$\rm O_{II}$] or UV fluxes \citep{ghost}.
The SFR estimation for the XRF 020903 host obtained using the $L_{\rm IR}$ method was one of the notable events that exhibited smaller values than the optical measurements, similar to GRB 031203 ($z=0.105$), GRB 050826 ($z=0.296$), GRB 051006 ($z=1.059$), and GRB 060814 ($z=1.923$).
This result indicates that the low dust content of the GRB host galaxies. Because the lower optical depth of the dust tends to increase the SFR conversion coefficient \citep[e.g.,][]{buat96, kennicutt98}, an empirical calibration of SFR/$L_{\rm IR}$ is required based on other physical parameters, such as stellar mass and metallicity.

\begin{figure*}
\epsscale{0.85}
\plotone{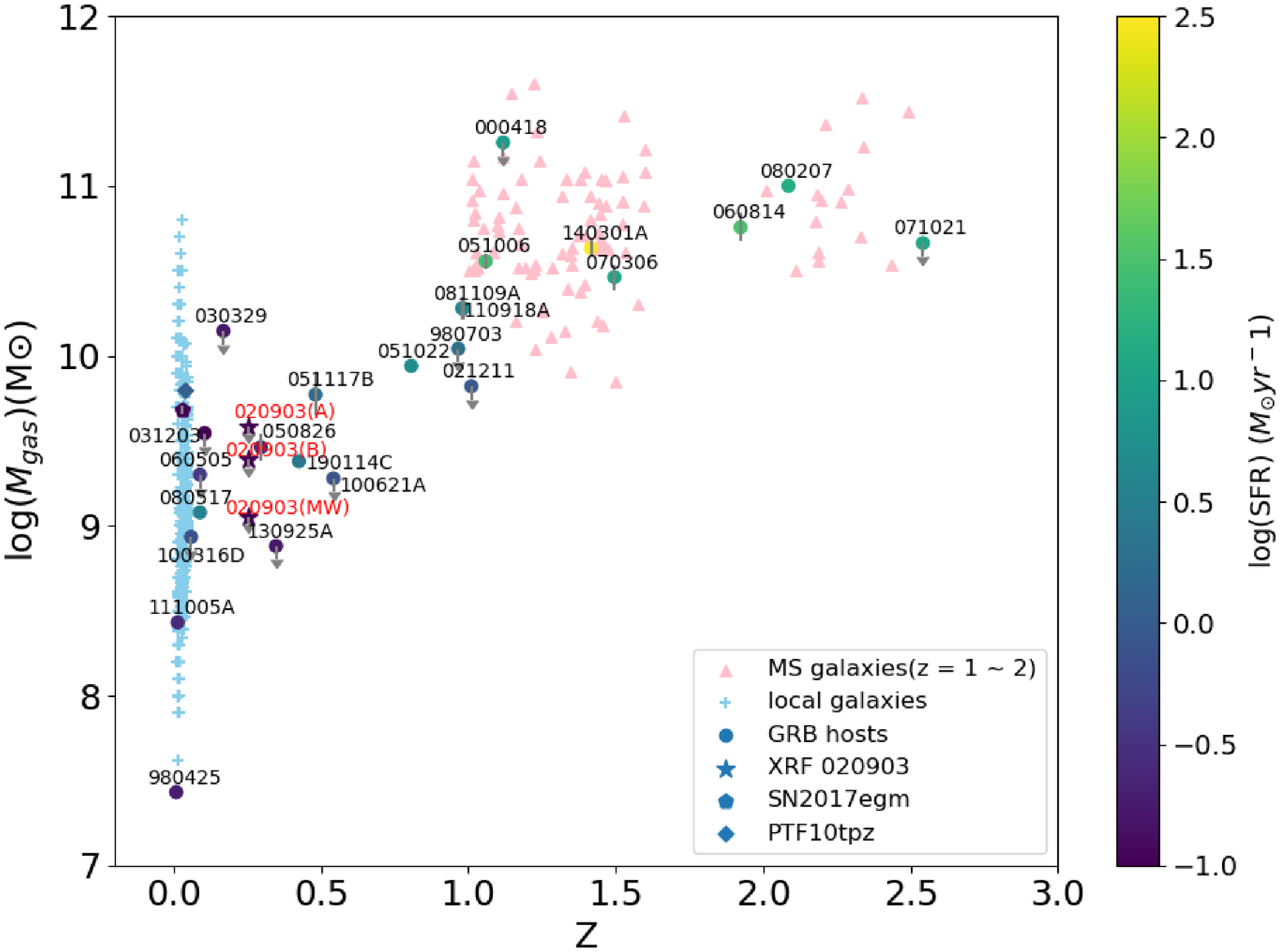}
\plotone{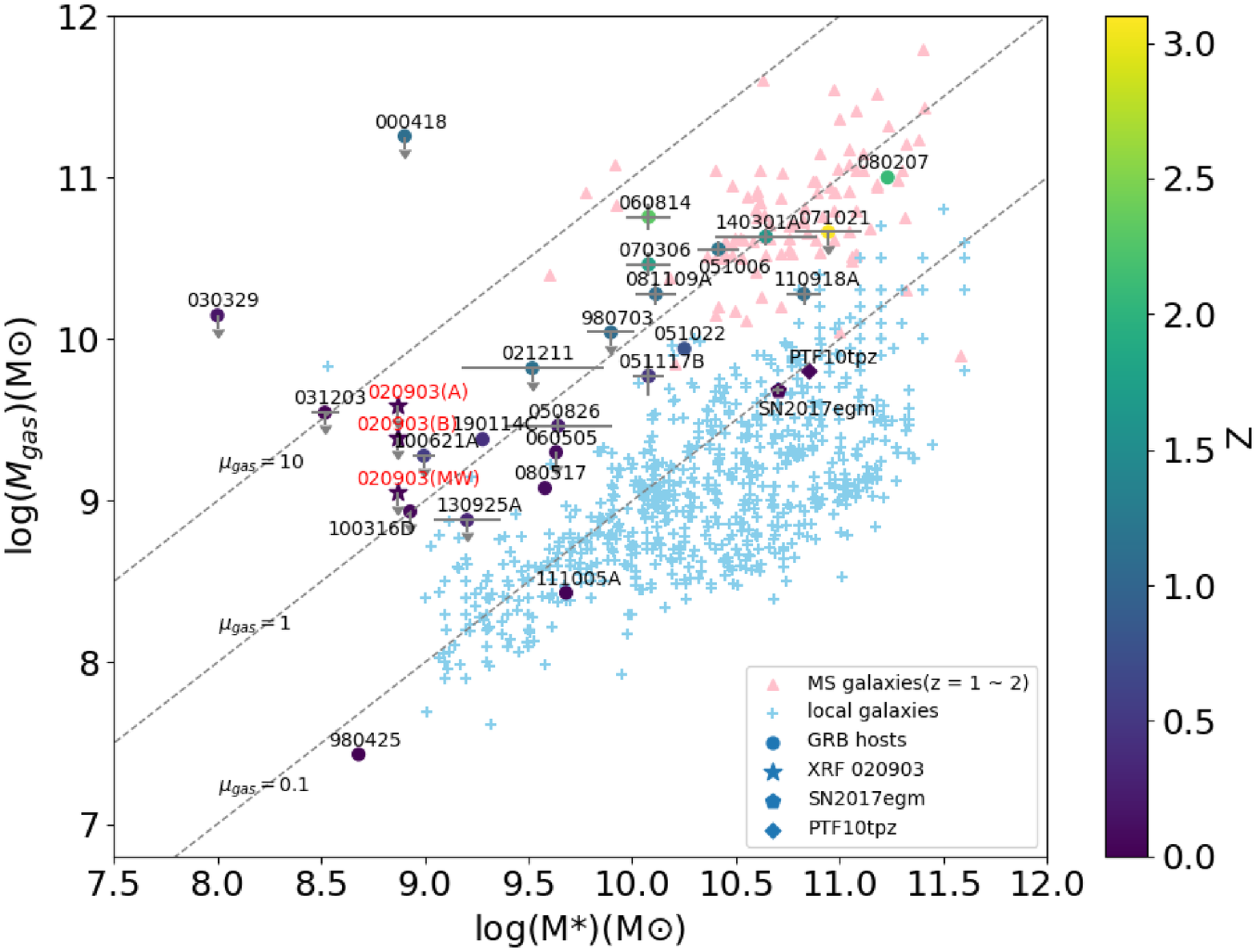}
\caption{(Top) Redshift dependency of molecular gas mass ($M_{\rm gas}$). Arrows represent 3$\sigma$ upper limits. (Bottom) Comparison between molecular gas mass ($M_{\rm gas}$) and stellar mass ($M_{*}$). Dashed lines show the molecular gas fractions ($\mu_{\rm gas} = M_{\rm gas}/M_{*}$) of 0.1, 1, and 10. 
Three upper limits of $M_{\rm gas}$ along to the metallicity-dependent CO-to-H$_{2}$ conversion factor are plotted for XRF 020903 (burst site as A, and remaining regions representative as B, and the value estimated with the galactic conversion factor as MW). For comparison, local galaxies \citep{both14, saintonge, tacconi18} and $z\sim1-2$ main-sequence galaxies are plotted. \citep{daddi10, magnelli12, tacconi13, seko16}.\label{mgas2mstar}}
\end{figure*}

\section{Limit of Luminosity of CO and Molecular Gas}
\subsection{Molecular Gas Mass}
The luminosity of CO was estimated based on the equation ($L'_{\rm CO} = 3.25 \times 10^{7}S_{\rm CO}\Delta\nu\nu_{obs}^{-2}D_{L}^{2}(1+z)^{-3}$) described by \citet{solomon05},  where $L'_{\rm CO}$ is in K km s$^{-1}$ pc$^{2}, S_{\rm CO}\Delta\nu$ is the velocity-integrated intensity in Jy km s$^{-1}$, $\nu_{\rm obs}$ is the observed line frequency in GHz, and  $D_{L}$ is the luminosity distance in Mpc.
Because the CO(1-0) line was not detected from the ALMA observation of the XRF 020903 host galaxy, we derived the upper limit of the CO luminosity by assuming a velocity width of 180 km s$^{-1}$ (Table \ref{tab}). The assumption of this velocity width is identical to that of nondetection samples reported by \citet{hatsukade20}.

The upper limit of the molecular gas mass of the XRF 020903 host galaxy was derived from $M_{gas}=\alpha_{\rm CO}L'_{\rm CO(1-0)}$, where $\alpha_{\rm CO}$ is the CO-to-H$_{2}$ conversion factor, including the contribution of the helium mass.
We estimated the conversion factor by considering the dependency on the gas-phase metallicity and increasing $\alpha_{CO}$ as metallicity is decreased (e.g., Wilson 1995; Arimoto et al. 1996; Kennicutt \& Evans 2012; Bolatto et al. 2013).  
The optical measurements indicate that all four regions of the XRF 020903 host galaxy are subsolar-metallicity \citep[12+log(O/H)=8.1-8.3;][]{thorp2018}. 
In the low-metallicity range, the conversion factor estimation becomes difficult, because the estimation methods have not yet been established. For example, the empirical relations between metallicity and $\alpha_{\rm CO}$ of Genzel et al. (2012) and Bolatto et al. (2013) differed. Therefore, we adopted the harmonic mean of the recipes of \citet{genzel12} and Bolatto et al. (2013), similar to \citet{hatsukade20}.
The converted upper limits of the gas mass were $M_{\rm gas}<3.9\times10^{9}$ $M_{\sun}$ with $\alpha_{\rm CO}=15$ $M_\sun$ (K km s$^{-1}$ pc$^2$) for region A and $M_{\rm gas}<2.5\times10^{9}$ $M_{\sun}$ with $\alpha_{\rm CO}=9.6$ (K km s$^{-1}$ pc$^2$) for regions B, C, and D. 
 %
\begin{figure*}
\epsscale{0.95}
\plotone{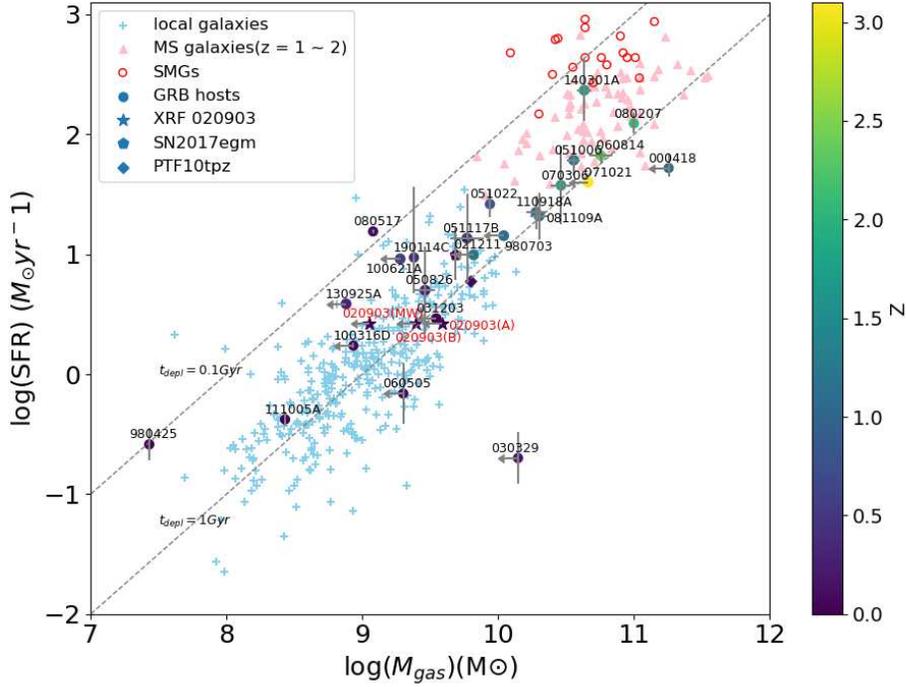}
\caption{Molecular gas mass ($M_{\rm gas}$) compared with SFR. Dashed lines indicate the gas depletion timescale of 0.1 and 1 Gyr(see Talbe \ref{tab}). Three upper limits of $M_{\rm gas}$ and metallicity-dependent CO-to-H$_{2}$ conversion factor are plotted for XRF 020903 (burst site as A, remaining regions representative as B, and the value estimated with the galactic conversion factor as MW). For comparison, local galaxies \citep{both14, saintonge}, $z\sim1-2$ main-sequence galaxies \citep{tacconi13, seko16}, and SMGs \citet{both13} are plotted.}
\label{sfrmgas}
\end{figure*}
The smaller upper limit of $M_{\rm gas}<1.1\times10^{9}$ was estimated by applying the galactic conversion factor of $\alpha_{\rm CO}$=4.4 \citep[K km s$^{-1}$ pc$^2$;][]{bolatto13b}.
We compared the molecular gas mass of the XRF 020903 with other GRB and SLSN samples reported by \citet{hatsukade20} as a function of redshift (Figure \ref{mgas2mstar}, top) and stellar mass (Figure \ref{mgas2mstar}, bottom). No violation was observed in the global trends of both the redshift and stellar mass.
Additionally, we compared the local star-forming galaxies \citep{both14, saintonge} and main-sequence galaxies at $z\sim1-2$ \citep{daddi10, magdis12, magnelli12, tacconi13, seko16}. 
The upper limit of $M_{\rm gas}$ was consistent with those of local galaxies, in addition to the redshift. A comparable sample in the same stellar mass range did not exist. 
The limits for the hosts of XRF 020903 is lower than the gas mass fractions for the most gas-rich galaxies and some GRB hosts.

\subsection{Molecular Gas Mass-SFR, Gas Fraction, and Depletion Timescale }
The gas surface density was correlated to the SFR surface density, which is known
as the Kennicutt-Schmidt relation \citep{schmidt59}. The integration over
the source region indicated a correlation between $M_{\rm gas}$ and the SFR.
Figure \ref{sfrmgas} shows the relationship between the GRB host galaxies and the
target. The SFR of XRF 020903 estimated using the $H_{\alpha}$ line as 2.65
$M_{\sun}$ yr$^{-1}$ \citep{ghost}, was used. Three upper limits of $M_{\rm gas}$
for XRF 020903 (burst site as A, remaining regions representative as B, and the value estimated with the galactic conversion factor as MW) 
were plotted, because the conversion factor of CO-to-H$_{2}$ depends on the metallicity, as described in $\S5.1$. 
The GRB samples were those of \citet{hatsukade20}, who employed the SED fitting method to estimate the SFRs.
The upper limit of the molecular gas mass of XRF 020903 indicates that the current event reflects the trend of GRB host galaxies.
Additionally, we compared the local star-forming galaxies \citep{both14, saintonge} and main-sequence galaxies at $z\sim1-2$ \citep{tacconi13, seko16}. The upper limit of XRF 020903 was consistent with that of the local star-forming galaxies. When we adopted the galactic conversion factor of $\alpha_{\rm CO}$=4.4 (K km s$^{-1}$ pc$^2$), the limit indicated that the XRF 020903 host galaxy was at the smallest $M_{\rm gas}$ end in the same SFR range.

\begin{figure*}
\epsscale{0.9}
\plotone{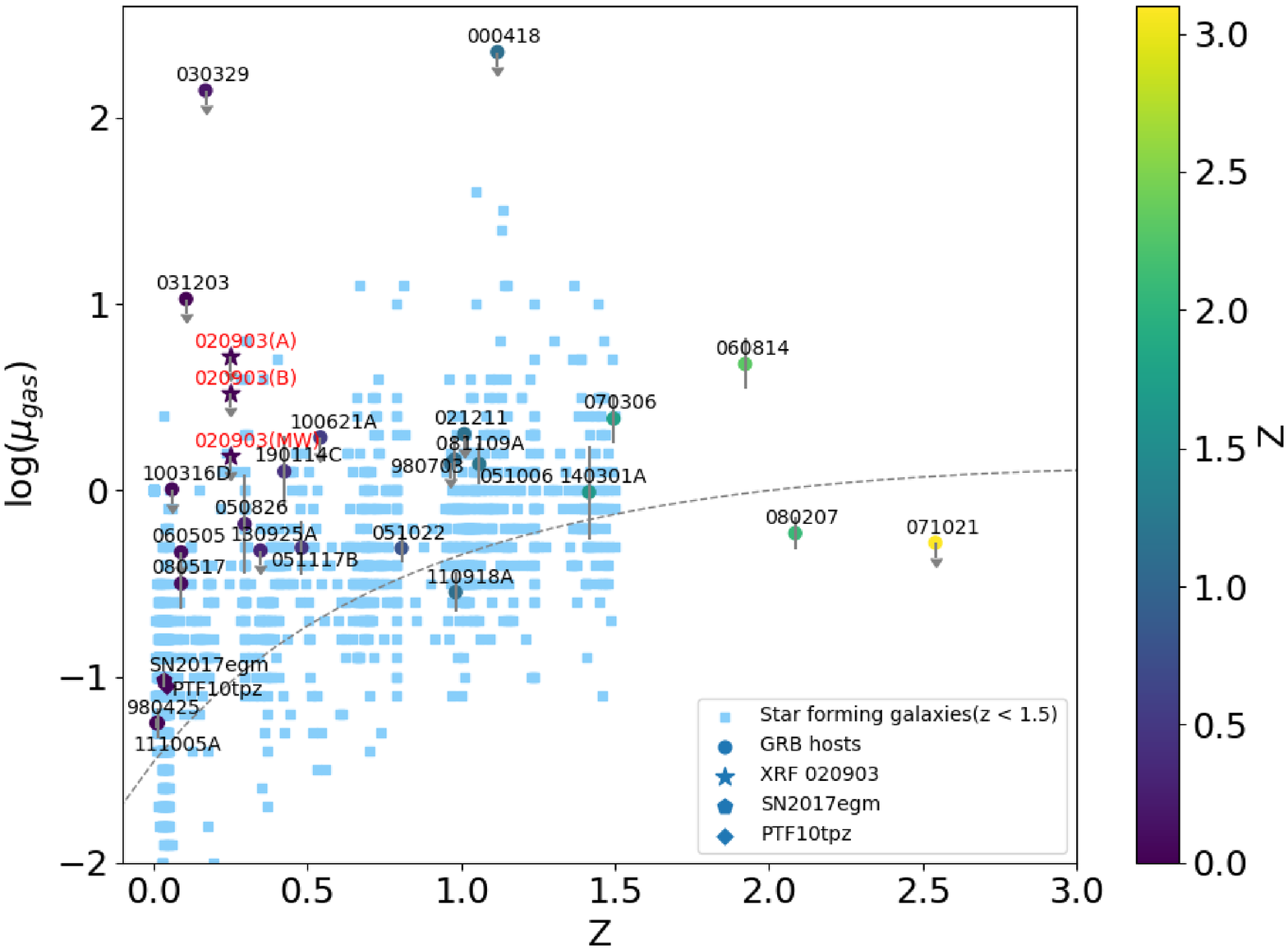}
\plotone{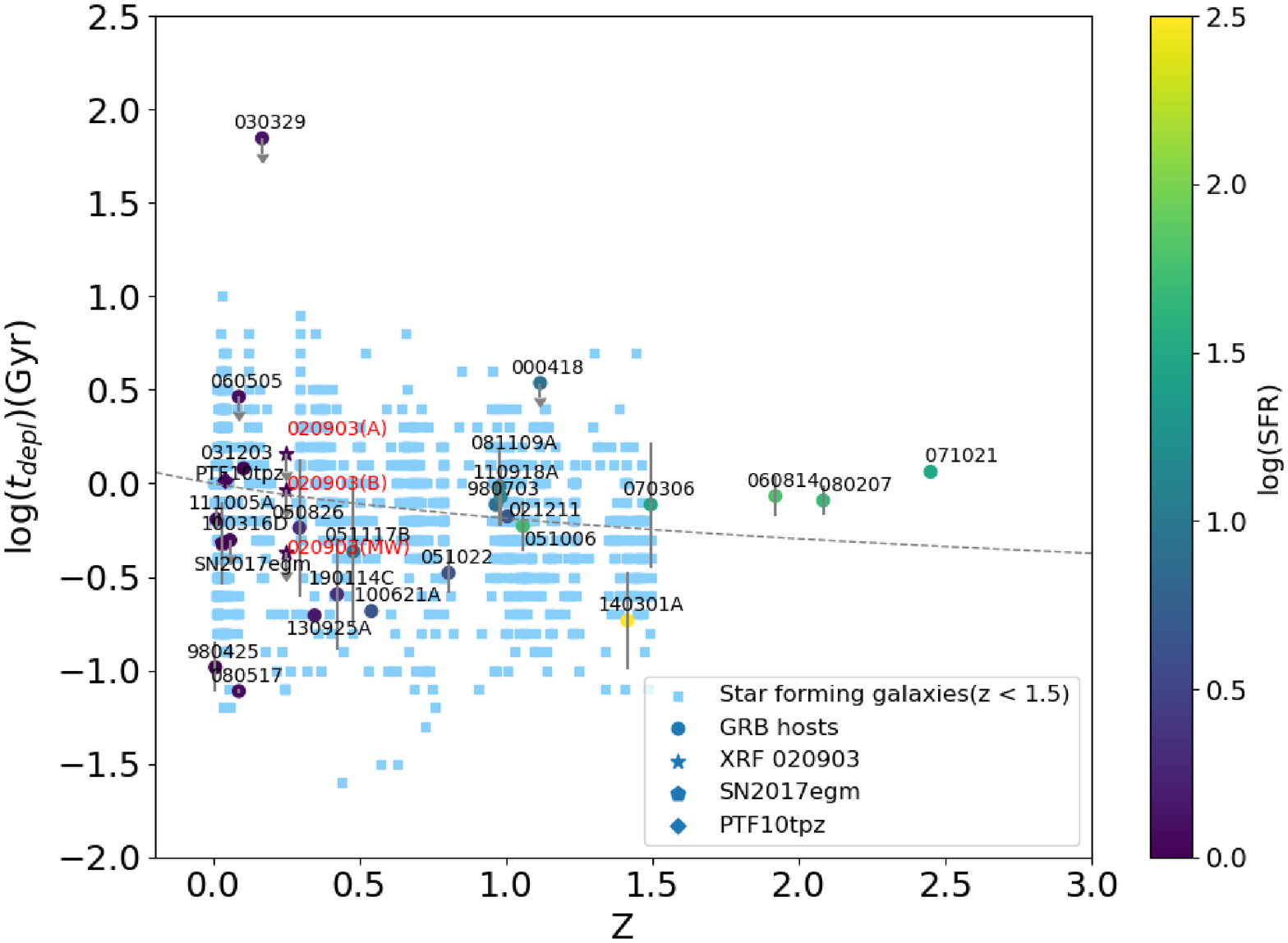}
\caption{Molecular gas fraction ($\mu_{\rm gas}$; top) and molecular gas depletion timescale ($t_{\rm depl}$; bottom) compared with the redshift.  Three upper limits and metalicity-dependent CO-to-H$_{2}$ conversion factor are plotted for XRF 020903 (burst site as A, remaining regions representative as B, and the value estimated with the galactic conversion factor as MW). The dashed line represents the best-fit line for the main-sequence galaxies derived by \citet{tacconi18}. Distribution of star-forming galaxies \citep{tacconi18} is plotted for comparison.
\label{tuz}}
\end{figure*}

The majority of the GRB host galaxies were located on the molecular gas depletion timescale ($t_{\rm depl}$=$M_{\rm gas}$/SFR) of $\sim$1 Gyr, whereas some hosts with lower redshift had a shorter gas depletion timescale (Figure \ref{sfrmgas}). 
\citet{hatsukade20} reported that GRB host galaxies tend to possess a higher molecular gas mass fraction ($\mu_{\rm gas}$)  and a shorter gas depletion timescale ($t_{\rm depl}$) than other star-forming galaxies at similar redshifts, particularly at $z<1$. Figure \ref{tuz} shows $\mu_{\rm gas}$ and $t_{\rm depl}$ as a function of redshift. 
The lines in Figure \ref{tuz} indicate the best-fit function of star-forming galaxies derived by \citet{tacconi18}. For comparison, the distribution of the star-forming galaxies \citep{saintonge, tacconi18} is shown. Although the upper limit of $\mu_{\rm gas}$ could not constrain the trend effectively, 
the limit of $t_{\rm depl}$ was consistent with the results of most of GRB samples (e.g., GRB 980703, GRB 021211, GRB 031203, GRB 050826, GRB 060814, GRB 070306, GRB 071021, GRB 081109A, GRB 110918A), and one of SLSN, PTF10tpz. 
As the upper limits were close to the best-fitted function of the star-forming galaxies, the depletion time, $t_{\rm depl}$ was consistent with or rather lower than the average of star-forming galaxies. 

\section{Summary}

Results of the ALMA dust continuum and CO observations of the XRF 020903 host galaxy were reported herein. The continuum observation provided the upper limit of the IR luminosity based on the template scaling method. The corresponding SFR $<2.2 M_{\sun}$ yr$^{-1}$ was consistent with the measurement using the $H_{\alpha}$ emission line or rather small.  
Based on the CO observation, $M_{\rm gas}<3.9\times10^{9}$ $M_{\sun}$ for the burst site by applying the metallicity-dependent CO-to-H$_{2}$ conversion factor. The gas mass matched with the global trend of the GRB host galaxies, in addition to the redshift and stellar mass.
As one of the potential common properties of GRB host galaxies, we confirmed that the gas depletion timescale is equivalent to those of the GRBs and SLSNs at the same redshift range (i.e., GRB 031203, GRB 050826, and PTF10tpz).
These results indicate the similar properties of the XRF 020903 host galaxy to those of GRB host galaxies and supports the identical origins of XRF 020903 and GRBs. 

\acknowledgments
This paper makes use of the following ALMA data: ADS/JAO.ALMA\#2015.1.01254 and \#2016.1.01333.S. ALMA is a partnership of ESO (representing its member states), NSF (USA), and NINS (Japan), together with NRC (Canada), MOST and ASIAA (Taiwan), and KASI (Republic of Korea), in cooperation with the Republic of Chile. The Joint ALMA Observatory is operated by ESO, AUI/NRAO, and NAOJ. This research has made use of the GHostS database (www.grbhosts.org), which is partly funded by Spitzer/NASA grant RSA agreement No. 1287913. Based on observations made with the NASA/ESA Hubble Space Telescope and obtained from the Hubble Legacy Archive, which is a collaboration between the Space Telescope Science Institute (STScI/NASA), the Space Telescope European Coordinating Facility (ST-ECF/ESA), and the Canadian Astronomy Data Centre (CADC/NRC/CSA).
This work is supported by the Ministry of Science and Technology of Taiwan grants MOST 105-2112-M-008-013-MY3 (Y.U.). 
%

\begin{longrotatetable}
\begin{deluxetable*}{ccccccccccc}
\tabletypesize{\scriptsize}
\tablenum{1}
\tablecaption{XRF 020903 and GRBs host galaxies samples\label{tab}}
\tablehead{
\colhead{Event} & 
\colhead{Redshift} &
\colhead{Frequency} & 
\colhead{Flux} &
\colhead{L$_{\rm IR}$} & 
\colhead{SFR} &
\colhead{CO Luminosity} & 
\colhead{Gas Mass} & 
\colhead{Reference} \\
            &        &  (GHz) &    (mJy)            &  ($L_{\odot}$)        &($M_{\odot}$ yr$^{-1}$)      &(K km s$^{-1} pc^{2}$)       & ($M_{\odot}$)             &             
}
\startdata      
GRB 980425  & 0.0085 & 106.8 & 0.77$\pm$0.26        & $3.53\times10^{9}$                &  0.60             & $2.3\times10^{6}$         & $2.7\times10^{7}$         & \citet{michalowski18} \\
GRB 980703  & 0.966  & 234.5 & $<$0.081             & $<1.75\times10^{11}$          & $<$29.7           & $<4.0\times10^{8}$        & $<1.1\times10^{10}$       & \citet{hatsukade20} \\
GRB 000418  & 1.1183 & $-$   & $-$                  & $-$                               &  $-$              &$<1.0\times10^{10}$        & $<1.8\times10^{11}$       & \citet{hatsukade11} \\ 
GRB 020819B & 1.9621 & 245.1 & 0.14$\pm$0.03        & $6.23\times10^{11}$           & 105.9             & $-$                       & $-$                       & \citet{hatsukade14} \\
XRF 020903  & 0.251  & 97.2  & $<$0.032             & $<1.28\times10^{11}$              & $<$21.7           & $<2.6\times10^{8}$        & $<1.1\times10^{9}$        & This work \\
            &        & 233.0 & $<$0.061             & $<1.30\times10^{10}$              & $<$2.21           & $-$                       & $-$                       & This work \\
GRB 021004  & 2.330  & 345.0 & $<$0.354             & $<3.13\times10^{11}$              & $<$53.1           & $-$                       & $-$                       & \citet{wang12} \\
GRB 021211  & 1.010  & 229.8 & $<$0.063             & $<1.5\times10^{11}$           & $<$25.8           & $<2.9\times10^{8}$        & $<6.6\times10^{9}$        & \citet{hatsukade20} \\
GRB 030329  & 0.1685 & $-$   & $-$                  & $-$                               & $-$               & $<6.9\times10^{8}$        & $<1.4\times10^{10}$       & \citet{endo07} \\
GRB 031203  & 0.105  & 312.8 & $<$2.410             & $<4.5\times10^{9}$            &   $<$0.8          & $<1.8\times10^{8}$        & $<3.5\times10^{9}$        & \citet{hatsukade20} \\
GRB 050401  & 2.900  & 343.5 & $<$0.67              & $<7.30\times10^{11}$              & $<$124.2          & $-$                       & $-$                       & This work \\
GRB 050709  & 0.1606 & 233.0 & $<$0.05              & $<5.97\times10^{9}$               & $<$1.0            & $-$                       & $-$                       & This work \\
GRB 050826  & 0.296  & 266.8 & $<$0.084             & $<2.4\times10^{10}$            & $<$4.1            & $(2.8\pm0.5)\times10^{8}$ & $(2.9\pm0.5)\times10^{9}$ & \citet{hatsukade20} \\
GRB 050915A & 2.527  & 343.5 & 0.37$\pm$0.13        & $4.09\times10^{11}$               & 69.5              & $-$                       & $-$                       & This work \\
GRB 051006  & 1.059  & 223.9 & 0.078$\pm$0.024      & $2.0\times10^{11}$            &   33.8            & $(2.7\pm0.3)\times10^{9}$ & $(3.6\pm0.4)\times10^{10}$& \citet{hatsukade20} \\
GRB 051022  & 0.806  & 256.1 & $<$1.6               & $<6.25\times10^{12}$              & $<$1062           & $4.2\times10^{8}$         & $8.7\times10^{9}$         & \citet{hatsukade14} \\
GRB 051117B & 0.481  & 225.3 & 0.069$\pm$0.020      & $8.0\times10^{10}$                &  13.7             & $(9.1\pm0.3)\times10^{8}$ & $(5.9\pm1.7)\times10^{9}$ & \citet{hatsukade20} \\
GRB 060505  & 0.089  & $-$   & $-$                  & $-$                               &  $-$              & $<4.2\times10^{8}$        & $<2.0\times10^{9}$        & \citet{michalowski18} \\
GRB 060814  & 1.923  & 343.5 & $<$0.063             & $<6.8\times10^{11}$           &  $<$115.1            & $(2.7\pm0.5)\times10^{9}$ & $(5.7\pm1.1)\times10^{10}$& \citet{hatsukade20} \\
GRB 070306  & 1.496  & 343.5 & $<$0.036             & $<5.7\times10^{11}$               &  $<$96.8          & $(2.2\pm0.4)\times10^{9}$ & $(2.9\pm0.5)\times10^{10}$& \citet{hatsukade20} \\
GRB 070521  & 2.087  & 680.0 & $<$3.12              & $<7.30\times10^{11}$              & $<$124.2          & $-$                       & $-$                       & \citet{hashimoto19} \\
GRB 070802  & 2.450  & 343.5 & 0.30$\pm$0.11        & $2.88\times10^{11}$               & 49.0              & $-$                       & $-$                       & This work \\
GRB 071021  & 2.542  & 133.6 & $<$0.042             & $<6.8\times10^{11}$           &  $<$115.1         & $<1.3\times10^{9}$        & $<4.6\times10^{10}$       & \citet{hatsukade20} \\
GRB 080207  & 2.086  & 142.5 & 0.11$\pm$0.03        & $2.46\times10^{12}$               &  417.5              & $1.7\times10^{10}$        & $1.0\times10^{11}$        & \citet{hatsukade19} \\
GRB 080517  & 0.0889 & $-$   & $-$                  & $-$                               &  $-$              & $1.5\times10^{8}$         & $1.2\times10^{9}$         & \citet{stanway15} \\
GRB 081109A & 0.979  & 233.0 & $<$0.087             & $<2.0\times10^{11}$           &   $<$33.8         & $(1.4\pm0.2)\times10^{9}$ & $(2.0\pm0.3)\times10^{10}$& \citet{hatsukade20} \\
GRB 081221  & 2.260  & 343.5 & 0.88$\pm$0.09        & $1.13\times10^{12}$               & 192.5             & $-$                       & $-$                       & This work \\
GRB 090423  & 8.230  & 222.0 & $<$0.036             & $<4.50\times10^{10}$              & $<$7.6            & $-$                       & $-$                       & \citet{berger14} \\
GRB 100316D & 0.059  & $-$   & $-$                  & $-$                               &  $-$              & $<1.4\times10^{8}$        & $<8.6\times10^{8}$        & \citet{michalowski18} \\
GRB 100621A & 0.542  & 224.3 &$<$0.066              & $<5.7\times10^{10}$           & $<$9.8            & $<1.4\times10^{8}$        & $<1.9\times10^{9}$        & \citet{hatsukade20} \\
GRB 110918A & 0.982  & 232.3 &0.097$\pm$0.020       & $2.4\times10^{11}$            &    41.4           & $(1.7\pm0.2)\times10^{9}$ & $(1.9\pm0.3)\times10^{10}$& \citet{hatsukade20} \\
GRB 111005A & 0.013  & $-$   & $-$                  & $-$                           &  $-$              & $4.4\times10^{7}$         & $2.7\times10^{8}$         & \citet{michalowski18} \\
GRB 111123A & 3.152  & 343.5 & 0.96$\pm$0.10        & $1.13\times10^{12}$           & 192.5             & $-$                       & $-$                       & This work \\
GRB 120624B & 2.197  & 343.5 & 0.35$\pm$0.12        & $3.68\times10^{11}$           & 62.5              & $-$                       & $-$                       & This work \\
GRB 130606A & 5.913  & 265.0 & $<$0.354             & $<5.69\times10^{11}$          & $<$96.8           & $-$                       & $-$                       & This work \\
GRB 130925A & 0.347  & 256.5 & $<$0.096             & $<3.8\times10^{10}$           & $<$6.5            & $<7.6\times10^{7}$       & $<7.6\times10^{8}$        & \citet{hatsukade20} \\
GRB 131030  & 1.293  & 345.0 & $<$0.12              & $<6.5\times10^{11}$           & $<$11.1           & $-$                       & $-$                       & \citet{huang17} \\
GRB 140301A & 1.416  & 143.2 & $<$0.039             & $<5.7\times10^{11}$           &  $<$96.8          & $(5.5\pm0.6)\times10^{9}$ & $(4.3\pm0.5)\times10^{10}$ & \citet{hatsukade20} \\
GRB 190114C & 0.425  & $-$   & $-$                  & $-$                           & $-$               & $1.54\times10^{8}$        & $2.4\times10^{9}$         & \citet{postigo20} \\
SN2017egm   & 0.031  & $-$   & $-$                  & $-$                           &  $-$              & $(1.1\pm0.1)\times10^{9}$ & $(4.8\pm0.3)\times10^{9}$ & \citet{hatsukade20egm} \\ 
PTF10tpz    & 0.040  & $-$   & $-$                  & $-$                           &   $-$             &$(1.45\pm0.03)\times10^{9}$& $6.3\times10^{9}$         & \citet{ptf10tpz} \\
\enddata
\end{deluxetable*}
\end{longrotatetable}


%

\facilities{ALMA, HST}


\software{CASA} \citep[v4.7.0-1;][]{casa}


\begin{thebibliography}{}

\bibitem[Alexander et al.(2017)]{alexander17} Alexander, K.~D., Berger, E., Fong, W., et al.\ 2017, \apjl, 848, L21
\bibitem[Amati et al.(2018)]{amati2018} Amati, L., O'Brien, P., G{\"o}tz, D., et al.\ 2018, Advances in Space Research, 62, 191. doi:10.1016/j.asr.2018.03.010
\bibitem[Arabsalmani et al.(2019)]{ptf10tpz} Arabsalmani, M., Roychowdhury, S., Renaud, F., et al.\ 2019, \apj, 882, 31. doi:10.3847/1538-4357/ab2897
\bibitem[Arabsalmani et al.(2020)]{co-grb980425} Arabsalmani, M., Renaud, F., Roychowdhury, S., et al.\ 2020, \apj, 899, 165. doi:10.3847/1538-4357/aba3c0
\bibitem[Arimoto et al.(1996)]{arimoto96} Arimoto, N., Sofue, Y., \& Tsujimoto, T.\ 1996, \pasj, 48, 275. doi:10.1093/pasj/48.2.275
\bibitem[Bellm et al.(2014)]{130925r1} Bellm, E.~C., Barri{\`e}re, N.~M., Bhalerao, V., et al.\ 2014, \apjl, 784, L19. doi:10.1088/2041-8205/784/2/L19
\bibitem[Berger et al.(2014)]{berger14} Berger, E., Zauderer, B.~A., Chary, R.-R., et al.\ 2014, \apj, 796, 96. doi:10.1088/0004-637X/796/2/96
\bibitem[Bersier et al.(2006)]{bersier06} Bersier, D., Fruchter, A.~S., Strolger, L.-G., et al.\ 2006, \apj, 643, 284. doi:10.1086/502640
\bibitem[Bolatto et al.(2013)]{bolatto13a} Bolatto, A.~D., Warren, S.~R., Leroy, A.~K., et al.\ 2013, \nat, 499, 450. doi:10.1038/nature12351
\bibitem[Bolatto et al.(2013)]{bolatto13b} Bolatto, A.~D., Wolfire, M., \& Leroy, A.~K.\ 2013, \araa, 51, 207. doi:10.1146/annurev-astro-082812-140944
\bibitem[Bothwell et al.(2013)]{both13} Bothwell, M.~S., Smail, I., Chapman, S.~C., et al.\ 2013, \mnras, 429, 3047. doi:10.1093/mnras/sts562
\bibitem[Bothwell et al.(2014)]{both14} Bothwell, M.~S., Wagg, J., Cicone, C., et al.\ 2014, \mnras, 445, 2599. doi:10.1093/mnras/stu1936
\bibitem[Buat \& Xu(1996)]{buat96} Buat, V. \& Xu, C.\ 1996, \aap, 306, 61
\bibitem[Chary \& Elbaz(2001)]{chary01} Chary, R., \& Elbaz, D.\ 2001, \apj, 556, 562 
\bibitem[Chen et al.(2020)]{chen20} Chen, W.~J., Urata, Y., Huang, K., et al.\ 2020, \apjl, 891, L15. doi:10.3847/2041-8213/ab76d4
\bibitem[Coppejans et al.(2018)]{coppejans18} Coppejans, D.~L., Margutti, R., Guidorzi, C., et al.\ 2018, \apj, 856, 56. doi:10.3847/1538-4357/aab36e
\bibitem[Covino et al.(2006)]{covino06} Covino, S., Malesani, D., Israel, G.~L., et al.\ 2006, \aap, 447, L5. doi:10.1051/0004-6361:200500228
\bibitem[Daddi et al.(2010)]{daddi10} Daddi, E., Bournaud, F., Walter, F., et al.\ 2010, \apj, 713, 686. doi:10.1088/0004-637X/713/1/686
\bibitem[de Ugarte Postigo et al.(2020)]{postigo20} de Ugarte Postigo, A., Th{\"o}ne, C.~C., Mart{\'\i}n, S., et al.\ 2020, \aap, 633, A68. doi:10.1051/0004-6361/201936668
\bibitem[Endo et al.(2007)]{endo07} Endo, A., Kohno, K., Hatsukade, B., et al.\ 2007, \apj, 659, 1431. doi:10.1086/512764
\bibitem[Evans et al.(2014)]{130925r3} Evans, P.~A., Willingale, R., Osborne, J.~P., et al.\ 2014, \mnras, 444, 250. doi:10.1093/mnras/stu1459
\bibitem[Fox et al.(2005)]{050709r3} Fox, D.~B., Frail, D.~A., Price, P.~A., et al.\ 2005, \nat, 437, 845. doi:10.1038/nature04189
\bibitem[Gal-Yam et al.(2006)]{galyam2006} Gal-Yam, A., Ofek, E.~O., Poznanski, D., et al.\ 2006, \apj, 639, 
\bibitem[Genzel et al.(2012)]{genzel12} Genzel, R., Tacconi, L.~J., Combes, F., et al.\ 2012, \apj, 746, 69. doi:10.1088/0004-637X/746/1/69
\bibitem[Godet et al.(2012)]{svom} Godet, O., Paul, J., Wei, J.~Y., et al.\ 2012, \procspie, 8443, 84431O. doi:10.1117/12.925171
\bibitem[Greiner et al.(2000)]{greiner2000} Greiner, J., Hartmann, D.~H., Voges, W., et al.\ 2000, \aap, 353, 998
\bibitem[Greiner et al.(2014)]{greiner14} Greiner, J., Yu, H.-F., Kr{\"u}hler, T., et al.\ 2014, \aap, 568, A75. doi:10.1051/0004-6361/201424250
\bibitem[Grindlay(1999)]{grindlay1999} Grindlay, J.~E.\ 1999, \apj, 510, 710
\bibitem[Haggard et al.(2017)]{haggard17} Haggard, D., Nynka, M., Ruan, J.~J., et al.\ 2017, \apjl, 848, L25 
\bibitem[Hammer et al.(2006)]{hammer2006} Hammer, F., Flores, H., Schaerer, D., et al.\ 2006, \aap, 454, 103. doi:10.1051/0004-6361:20064823
\bibitem[Han et al.(2010)]{han2010} Han, X.~H., Hammer, F., Liang, Y.~C., et al.\ 2010, \aap, 514, A24. doi:10.1051/0004-6361/200912475
\bibitem[Hashimoto et al.(2019)]{hashimoto19} Hashimoto, T., Hatsukade, B., Goto, T., et al.\ 2019, \mnras, 488, 5029. doi:10.1093/mnras/stz2034
\bibitem[Hatsukade et al.(2011)]{hatsukade11} Hatsukade, B., Kohno, K., Endo, A., et al.\ 2011, \apj, 738, 33. doi:10.1088/0004-637X/738/1/33
\bibitem[Hatsukade et al.(2014)]{hatsukade14} Hatsukade, B., Ohta, K., Endo, A., et al.\ 2014, \nat, 510, 247. doi:10.1038/nature13325
\bibitem[Hatsukade et al.(2019)]{hatsukade19} Hatsukade, B., Hashimoto, T., Kohno, K., et al.\ 2019, \apj, 876, 91. doi:10.3847/1538-4357/ab1649
\bibitem[Hatsukade et al.(2020a)]{hatsukade20} Hatsukade, B., Ohta, K., Hashimoto, T., et al.\ 2020, \apj, 892, 42. doi:10.3847/1538-4357/ab7992
\bibitem[Hatsukade et al.(2020b)]{hatsukade20egm} Hatsukade, B., Morokuma-Matsui, K., Hayashi, M., et al.\ 2020, \pasj, 72, L6. doi:10.1093/pasj/psaa052
\bibitem[Heise et al.(2001)]{heise2001} Heise, J., Zand, J.~I., Kippen, R.~M., et al.\ 2001, Gamma-ray Bursts in the Afterglow Era, 16. doi:10.1007/10853853\_4
\bibitem[Heise(2003)]{heise2003} Heise, J.\ 2003, Gamma-Ray Burst and Afterglow Astronomy 2001: A Workshop Celebrating the First Year of the HETE Mission, 662, 229 
\bibitem[Hjorth et al.(2005)]{050709r1} Hjorth, J., Watson, D., Fynbo, J.~P.~U., et al.\ 2005, \nat, 437, 859. doi:10.1038/nature04174
\bibitem[Huang et al.(2002)]{huang2002} Huang, Y.~F., Dai, Z.~G., \& Lu, T.\ 2002, \mnras, 332, 735 
\bibitem[Huang et al.(2017)]{huang17} Huang, K., Urata, Y., Takahashi, S., et al.\ 2017, \pasj, 69, 20
\bibitem[Huang et al.(2019)]{huang19} Huang, K., Shimoda, J., Urata, Y., et al.\ 2019, \apjl, 878, L25. doi:10.3847/2041-8213/ab23fd
\bibitem[Huang et al.(2020)]{huang2020} Huang, Y.-J., Urata, Y., Huang, K., et al.\ 2020, \apj, 897, 69. doi:10.3847/1538-4357/ab8f9a
\bibitem[Ioka, \& Nakamura(2018)]{ioka18} Ioka, K., \& Nakamura, T.\ 2018, Progress of Theoretical and Experimental Physics, 2018, 043E02
\bibitem[Izzo et al.(2019)]{izzo19} Izzo, L., de Ugarte Postigo, A., Maeda, K., et al.\ 2019, \nat, 565, 324. doi:10.1038/s41586-018-0826-3
\bibitem[Izzo et al.(2020)]{izzo20} Izzo, L., Auchettl, K., Hjorth, J., et al.\ 2020, \aap, 639, L11. doi:10.1051/0004-6361/202038152
\bibitem[Jin et al.(2018)]{jin18} Jin, Z.-P., Li, X., Wang, H., et al.\ 2018, \apj, 857, 128
\bibitem[Kathirgamaraju et al.(2018)]{kathirgamaraju18} Kathirgamaraju, A., Barniol Duran, R., \& Giannios, D.\ 2018, \mnras, 473, L121
\bibitem[Kennicutt \& Evans(2012)]{kennicutt12} Kennicutt, R.~C. \& Evans, N.~J.\ 2012, \araa, 50, 531. doi:10.1146/annurev-astro-081811-125610
\bibitem[Kennicutt(1998)]{kennicutt98} Kennicutt, R.~C., Jr.\ 1998, \araa, 36, 189 
\bibitem[Kippen et al.(2002)]{kippen2002} Kippen, R.~M., Woods, P.~M., Heise, J., et al.\ 2002, APS April Meeting Abstracts
\bibitem[Lamb \& Kobayashi(2018)]{2018MNRAS.478..733L} Lamb, G.~P., \& Kobayashi, S.\ 2018, \mnras, 478, 733
\bibitem[Lamb et al.(2005)]{lamb2005} Lamb, D.~Q., Donaghy, T.~Q., \& Graziani, C.\ 2005, \apj, 620, 355. doi:10.1086/426099
\bibitem[Lamb et al.(2019)]{2019ApJ...870L..15L} Lamb, G.~P., Lyman, J.~D., Levan, A.~J., et al.\ 2019, \apjl, 870, L15
\bibitem[Law et al.(2018)]{Law18} Law, C.~J., Gaensler, B.~M., Metzger, B.~D., et al.\ 2018, \apjl, 866, L22
\bibitem[Lazzati et al.(2017)]{lazzati17} Lazzati, D., L{\'o}pez-C{\'a}mara, D., Cantiello, M., et al.\ 2017, \apjl, 848, L6
\bibitem[Levesque et al.(2010)]{levesque2010} Levesque, E.~M., Berger, E., Kewley, L.~J., et al.\ 2010, \aj, 139, 694. doi:10.1088/0004-6256/139/2/694
\bibitem[Levinson et al.(2002)]{levinson2002} Levinson, A., Ofek, E.~O., Waxman, E., et al.\ 2002, \apj, 576, 923
\bibitem[Lyman et al.(2018)]{2018NatAs...2..751L} Lyman, J.~D., Lamb, G.~P., Levan, A.~J., et al.\ 2018, NatAs, 2, 751
\bibitem[Magdis et al.(2012)]{magdis12} Magdis, G.~E., Daddi, E., B{\'e}thermin, M., et al.\ 2012, \apj, 760, 6. doi:10.1088/0004-637X/760/1/6
\bibitem[Magnelli et al.(2012)]{magnelli12} Magnelli, B., Saintonge, A., Lutz, D., et al.\ 2012, \aap, 548, A22. doi:10.1051/0004-6361/201220074
\bibitem[Marcote et al.(2019)]{Marcote19} Marcote, B., Nimmo, K., Salafia, O.~S., et al.\ 2019, \apjl, 876, L14
\bibitem[McMullin et al.(2007)]{casa} McMullin, J.~P., Waters, B., Schiebel, D., Young, W., \& Golap, K.\ 2007, adass, 376, 127
\bibitem[Micha{\l}owski et al.(2018)]{michalowski18} Micha{\l}owski, M.~J., Karska, A., Rizzo, J.~R., et al.\ 2018, \aap, 617, A143. doi:10.1051/0004-6361/201833250
\bibitem[Murguia-Berthier et al.(2017)]{murguia17} Murguia-Berthier, A., Ramirez-Ruiz, E., Kilpatrick, C.~D., et al.\ 2017, \apjl, 848, L34
\bibitem[Nicholl et al.(2017)]{nicholl17} Nicholl, M., Berger, E., Margutti, R., et al.\ 2017, \apjl, 845, L8. doi:10.3847/2041-8213/aa82b1
\bibitem[Pastorello et al.(2010)]{slsn} Pastorello, A., Smartt, S.~J., Botticella, M.~T., et al.\ 2010, \apjl, 724, L16. doi:10.1088/2041-8205/724/1/L16
\bibitem[Perley et al.(2017)]{perley17} Perley, D.~A., Kr{\"u}hler, T., Schady, P., et al.\ 2017, \mnras, 465, L89. doi:10.1093/mnrasl/slw221
\bibitem[Piro et al.(2014)]{130925r2} Piro, L., Troja, E., Gendre, B., et al.\ 2014, \apjl, 790, L15. doi:10.1088/2041-8205/790/2/L15
\bibitem[Racusin et al.(2009)]{racusin09} Racusin, J.~L., Liang, E.~W., Burrows, D.~N., et al.\ 2009, \apj, 698, 43
\bibitem[Rau et al.(2006)]{rau2006} Rau, A., Greiner, J., \& Schwarz, R.\ 2006, \aap, 449, 79
\bibitem[Saintonge et al.(2017)]{saintonge} Saintonge, A., Catinella, B., Tacconi, L.~J., et al.\ 2017, \apjs, 233, 22. doi:10.3847/1538-4365/aa97e0
\bibitem[Sakamoto et al.(2004)]{sakamoto2004} Sakamoto, T., Lamb, D.~Q., Graziani, C., et al.\ 2004, \apj, 602, 875 
\bibitem[Sakamoto et al.(2005)]{sakamoto2005} Sakamoto, T., Lamb, D.~Q., Kawai, N., et al.\ 2005, \apj, 629, 311. doi:10.1086/431235
\bibitem[Savaglio et al.(2009)]{ghost} Savaglio, S., Glazebrook, K., \& Le Borgne, D.\ 2009, \apj, 691, 182. doi:10.1088/0004-637X/691/1/182
\bibitem[Schmidt(1959)]{schmidt59} Schmidt, M.\ 1959, \apj, 129, 243. doi:10.1086/146614
\bibitem[Seko et al.(2016)]{seko16} Seko, A., Ohta, K., Yabe, K., et al.\ 2016, \apj, 819, 82. doi:10.3847/0004-637X/819/1/82
\bibitem[Shirasaki et al.(2003)]{shirasaki2003} Shirasaki, Y., Kawai, N., Yoshida, A., et al.\ 2003, \pasj, 55, 1033. doi:10.1093/pasj/55.5.1033
\bibitem[Soderberg et al.(2004)]{soderberg2004} Soderberg, A.~M., Kulkarni, S.~R., Berger, E., et al.\ 2004, \apj, 606, 994 
\bibitem[Soderberg et al.(2005)]{soderberg2005} Soderberg, A.~M., Kulkarni, S.~R., Fox, D.~B., et al.\ 2005, \apj, 627, 877. doi:10.1086/430405
\bibitem[Solomon and Vanden Bout(2005)]{solomon05} Solomon, P.~M. \& Vanden Bout, P.~A.\ 2005, \araa, 43, 677. doi:10.1146/annurev.astro.43.051804.102221
\bibitem[Stanway et al.(2015)]{stanway15} Stanway, E.~R., Levan, A.~J., Tanvir, N.~R., et al.\ 2015, \apjl, 798, L7. doi:10.1088/2041-8205/798/1/L7
\bibitem[Tacconi et al.(2013)]{tacconi13} Tacconi, L.~J., Neri, R., Genzel, R., et al.\ 2013, \apj, 768, 74. doi:10.1088/0004-637X/768/1/74
\bibitem[Tacconi et al.(2018)]{tacconi18} Tacconi, L.~J., Genzel, R., Saintonge, A., et al.\ 2018, \apj, 853, 179. doi:10.3847/1538-4357/aaa4b4
\bibitem[Tanvir et al.(2009)]{tanvir09} Tanvir, N.~R., Fox, D.~B., Levan, A.~J., et al.\ 2009, \nat, 461, 1254. doi:10.1038/nature08459
\bibitem[Thorp \& Levesque(2018)]{thorp2018} Thorp, M.~D. \& Levesque, E.~M.\ 2018, \apj, 856, 36. doi:10.3847/1538-4357/aab093
\bibitem[Totani et al.(2014)]{totani14} Totani, T., Aoki, K., Hattori, T., et al.\ 2014, \pasj, 66, 63. doi:10.1093/pasj/psu032
\bibitem[Totani et al.(2016)]{totani16} Totani, T., Aoki, K., Hattori, T., et al.\ 2016, \pasj, 68, 15. doi:10.1093/pasj/psv123
\bibitem[Troja et al.(2018)]{troja18} Troja, E., Piro, L., Ryan, G., et al.\ 2018, \mnras, 478, L18
\bibitem[Troja et al.(2019)]{troja19} Troja, E., van Eerten, H., Ryan, G., et al.\ 2019, \mnras, 489, 1919
\bibitem[Urata et al.(2015)]{urata2015} Urata, Y., Huang, K., Yamazaki, R., \& Sakamoto, T.\ 2015, \apj, 806, 222 
\bibitem[Urata et al.(2019)]{urata19} Urata, Y., Toma, K., Huang, K., et al.\ 2019, \apjl, 884, L58
\bibitem[Villasenor et al.(2005)]{050709r2} Villasenor, J.~S., Lamb, D.~Q., Ricker, G.~R., et al.\ 2005, \nat, 437, 855. doi:10.1038/nature04213
\bibitem[Wang et al.(2012)]{wang12} Wang, W.-H., Chen, H.-W., \& Huang, K.-Y.\ 2012, \apjl, 761, L32. doi:10.1088/2041-8205/761/2/L32
\bibitem[Wheeler et al.(2017)]{wheeler17} Wheeler, J.~C., Chatzopoulos, E., Vink{\'o}, J., et al.\ 2017, \apjl, 851, L14. doi:10.3847/2041-8213/aa9d84
\bibitem[Wilson(1995)]{wilson95} Wilson, C.~D.\ 1995, \apjl, 448, L97. doi:10.1086/309615
\bibitem[Yamazaki et al.(2002)]{yamazaki2002} Yamazaki, R., Ioka,  K., \& Nakamura, T.\ 2002, \apjl, 571, L31 
\bibitem[Yasuda et al.(2017)]{yasuda17} Yasuda, T., Urata, Y., Enomoto, J., et al.\ 2017, \mnras, 466, 4558. doi:10.1093/mnras/stw3130
\bibitem[Yonetoku et al.(2020)]{yonetoku2020} Yonetoku, D., Mihara, T., Doi, A., et al.\ 2020, \procspie, 11444, 114442Z. doi:10.1117/12.2560603
\bibitem[Zhang et al.(2004)]{zhang2004} Zhang, B., Dai, X., Lloyd-Ronning, N.~M., \& M{\'e}sz{\'a}ros, P.\ 2004, \apjl, 601, L119 



\end{thebibliography}
\end{document}